\documentclass[superscriptaddress,twocolumn,prb]{revtex4-2}
\usepackage{bm}
\usepackage{amssymb}
\usepackage{amsmath}
\usepackage{float}
\usepackage[]{changes}
\usepackage{graphicx}
\usepackage[colorlinks=True,linkcolor=blue,anchorcolor=red,citecolor=blue,CJKbookmarks=true,urlcolor=blue]{hyperref}

\begin{document}

\title{Interplay between defects and the non-Hermitian skin effect}
\author{Yin Huang} \email{yhuan15@csu.edu.cn}
\affiliation{School of Physics, Central South University, Changsha, 410083, China}
\author{Wenna Zhang}
\affiliation{School of Physics, Central South University, Changsha, 410083, China}
\author{Yu Zhou}
\affiliation{Department of Physics, Jiangsu University of Science and Technology, Jiangsu 212003, China}
\author{Yuecheng Shen} \email{ycshen@lps.ecnu.edu.cn}
\affiliation{State Key Laboratory of Precision Spectroscopy, East China Normal University, Shanghai, 200241, China}
\author{Georgios Veronis}
\affiliation{School of Electrical Engineering and Computer Science, Louisiana State University, Baton Rouge, Louisiana 70803, USA}	
\affiliation{Center for Computation and Technology, Louisiana State University, Baton Rouge, Louisiana 70803, USA}
\author{Wenchen Luo}\email{luo.wenchen@csu.edu.cn}
\affiliation{School of Physics, Central South University, Changsha, 410083, China}

	\date{\today}
	
 \begin{abstract}
The non-Hermitian skin effect (NHSE) is an intriguing phenomenon in which an extensive number of bulk eigenstates localize at the boundaries of a non-Hermitian system with non-reciprocal hoppings. Here we study the interplay
of this effect and a defect in non-reciprocal one-dimensional lattices. We show that the interplay of the NHSE and defects is size-dependent. We demonstrate a novel class of hybrid skin-defect states in finite-size systems resulting from the coupling between the skin and defect states. Next, we consider a single defect in a topologically nontrivial lattice with time-reversal symmetry based on the non-reciprocal Su-Schrieffer-Heeger configuration. We unveil how topologically nontrivial defect states and non-Hermiticity interplay by competing with each other, exhibiting a transition from topologically nontrivial defect states to skin states. In addition, we show that decreasing the defect strength can result in a transition from trivial defect states to skin states. Our work promotes the understanding of the interplay between defects and the NHSE, and especially the importance of the energy spectra of the system with the defect under periodic boundary conditions.
\end{abstract}

\maketitle
\section{Introduction}
Non-Hermitian physics has flourished over the past few years revealing numerous unique phenomena of which
the counterparts cannot be found in Hermitian systems \cite{peng2014parity,ramezani2010unidirectional,chang2014parity,chen2016topological,zhou2018observation,shen2018quantum,papaj2019nodal,yoshida2018non}. One of these intriguing phenomena is the non-Hermitian skin effect (NHSE) \cite{yao2018edge, PhysRevX.9.041015,zhang2022review, Zhang2021Correspondence,kunst2018biorthogonal,Li2022Gain}, characterized by the localization of eigenstates at boundaries. This edge effect leads to the breakdown of the bulk-boundary correspondence associated with the point gap topology. To date, the NHSE has been extensively investigated both theoretically and experimentally in various systems \cite{weidemann2020topological,wang2021generating,zhang2021acoustic,xiao2020non,zhang2021observation,gu2022transient,wang2022non,li2020critical,helbig2020generalized,zhu2022anomalous,franca2022non,Longhi2022Self,Li2023band}.

Defect bound states exhibit localization behavior and are important in affecting the transport properties of the system. They are candidates for use in quantum computing \cite{Elliott2015,Aasen2016}, sensing \cite{Wolfowicz2021, Zhang2009}, spectroscopy \cite{Grasser2010,Kaftelen2012}, and switching applications \cite{Ren2004}. In non-Hermitian systems, the defect states exist in non-Hermitian flatbands \cite{Qi2018}, acquire topological protection \cite{Malzard2015}, and exhibit interesting topological effects \cite{Stegmaier2021}. Recently, the physics of impurities in non-reciprocal lattices has been investigated \cite{Liu2021,Schindler2021,Sun2021geometry,Roccati2021}. However, the impurities or defects in these non-reciprocal systems are introduced to induce boundaries in periodic systems. A strong impurity or defect behaves similarly to an open boundary condition (OBC) and effectively leads to the NHSE in a non-reciprocal lattice. Some theoretical efforts have also been made recently to explore the interplay between the NHSE and external magnetic fields \cite{lu2021magnetic, Shao2022Cyclotron, Li2023enhancement} or Anderson localization \cite{Jiang2019}. Defects tend to localize states in the bulk of the system in contrast to the NHSE. Thus, two natural questions arise: What is the physical picture when a defect is introduced into the middle of a non-reciprocal lattice under OBC, and what is the fate of defect states in the presence of the NHSE?

In this paper, we first study the interplay of the NHSE and defects in the non-reciprocal one-dimensional Hatano-Nelson (HN) model. We show that this interplay is size-dependent. We find a novel class of hybrid skin-defect states in finite-size systems formed by the interaction between the skin and defect states. We also find that the transition between hybrid skin-defect states and skin states depends on the size and the non-reciprocity strength of the system. In addition, we investigate a single defect in a topologically nontrivial lattice with time-reversal symmetry based on the non-reciprocal Su-Schrieffer-Heeger (SSH) model. We also demonstrate how non-reciprocity leads to a transition from topologically nontrivial defect states to skin states. Decreasing the defect strength can also result in a transition from trivial defect states to skin states.

\section{Single defect in a non-reciprocal HN-model-based system}

\begin{figure}[h]
\centering
\includegraphics[width=\columnwidth]{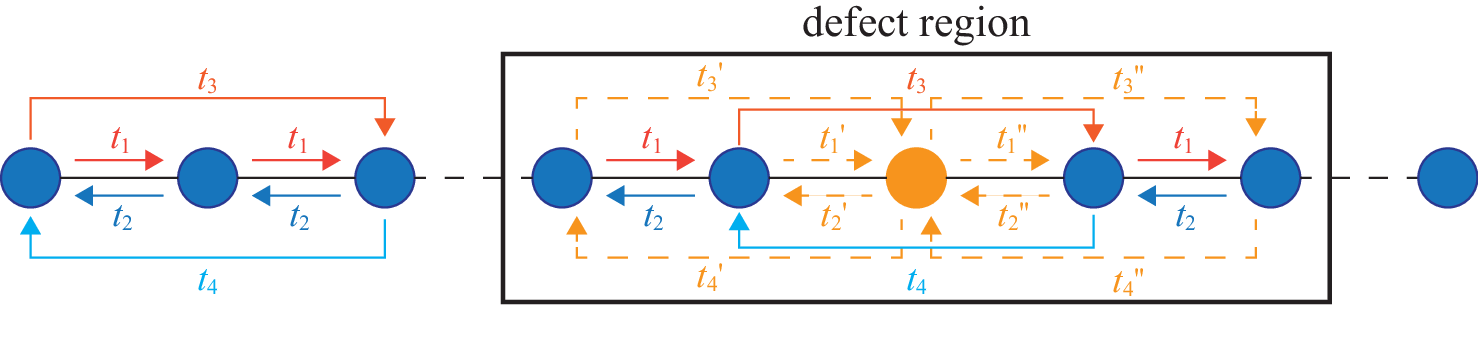}
\caption{Schematic of a non-Hermitian one-dimensional HN model with a defect.}
\label{fig 1}
\end{figure}

In non-Hermitian systems, eigenstates localize at the boundaries, exhibiting the NHSE, with the topological property that the OBC spectra are entirely enclosed by the periodic boundary condition (PBC) spectra \cite{yao2018edge}. We first consider a non-Hermitian HN-model-based system \cite{Hatano1996} with a defect, illustrated in Fig. \ref{fig 1}.
The Hamiltonian of such a non-reciprocal system can be written as
\begin{equation} \label{modelnh}
H_{HN}=H_{HN}^0 + H_{HN}^d,
\end{equation}
where $H_{HN}^0$ is the Hamiltonian without the defect region containing the left and right branches of the chain, while the term $H_{HN}^d$ describes the defect region.
The Hamiltonian without the defect is
\begin{eqnarray} \label{modelnh1}
H_{HN}^0&=&\sum_{n =1}^{N_d-2} (t_2 c_{n}^{\dag}c_{n+1}+t_1 c_{n+1}^{\dag}c_{n}) \notag \\
&+& \sum_{n=N_d+1}^{N-1}(t_2c_{n}^{\dag}c_{n+1}+t_1c_{n+1}^{\dag}c_{n})
\notag \\
&+& \sum_{n=1}^{N_d-3} (t_4c_{n}^{\dag}c_{n+2}+t_3c_{n+2}^{\dag}c_{n}) \notag \\
&+& \sum_{n=N_d+1}^{N-2}(t_4c_{n}^{\dag}c_{n+2}+t_3c_{n+2}^{\dag}c_{n}),
\end{eqnarray}
where $c_{n}$ and $c_{n}^{\dag}$ are the annihilation and creation operators, respectively, at site $n$ of the lattice; $N$ is the overall number of sites and $N_d$ is the position of the defect; $t_1$ and $t_2$ are non-reciprocal hopping terms representing the nearest-neighbor (NN) hopping; $t_3$ and $t_4$ are non-reciprocal hopping terms representing the next-nearest-neighbor (NNN) hopping. The second term $H_{HN}^d$ in Eq. (\ref{modelnh}) corresponds to the Hamiltonian of the defect region, which is given by
\begin{eqnarray} \label{modelnh2}
H_{HN}^d &=& ({t'_2}c_{N_d-1}^{\dag}c_{N_d}+{t'_1}c_{N_d}^{\dag}c_{N_d-1})  \notag \\
&+&({t''_2}c_{N_d}^{\dag}c_{N_d+1}+{t''_1}c_{N_d+1}^{\dag}c_{N_d})  \notag \\
&+&({t_4}c_{N_d-1}^{\dag}c_{N_d+1}+{t_3}c_{N_d+1}^{\dag}c_{N_d-1}) \notag \\
&+&({t'_4}c_{N_d-2}^{\dag}c_{N_d}+{t'_3}c_{N_d}^{\dag}c_{N_d-2})  \notag \\
&+&({t''_4}c_{N_d}^{\dag}c_{N_d+2}+{t''_3}c_{N_d+2}^{\dag}c_{N_d}),
\end{eqnarray}
where $t'_1$, $t'_2$, $t''_1$, and $t''_2$ are NN hopping terms in the defect region; $t'_3$, $t'_4$, $t''_3$, and $t''_4$ are NNN hopping terms in the defect region.

\begin{figure}[h]
	\centering
	\includegraphics[width=\columnwidth]{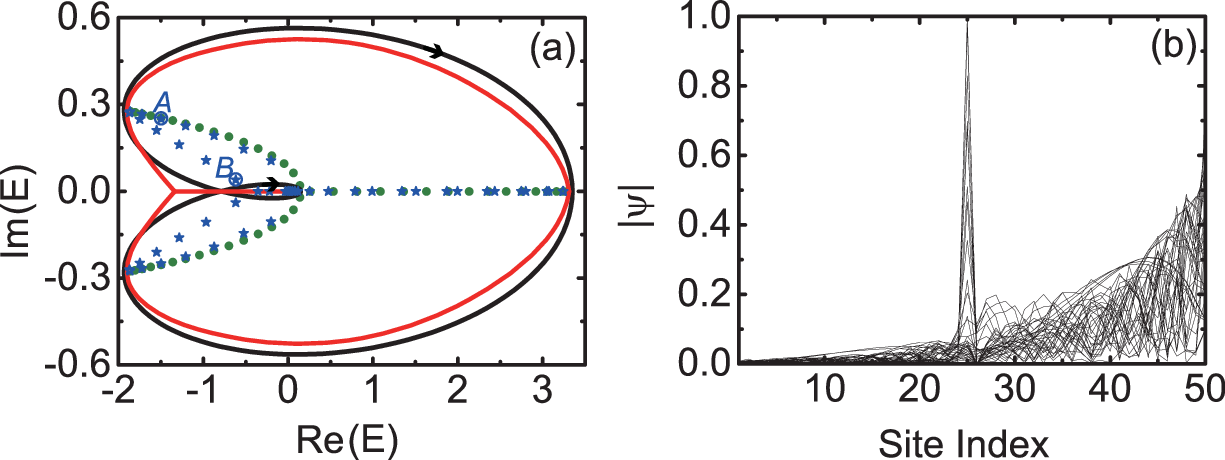}
	\caption{The eigenstates and the energy spectra of the one-dimensional non-Hermitian HN model including a defect with $t_1=1$, $t_2=0.6$, $t_3=1$, $t_4=0.75$, $t'_4 = t''_3=t'_2 = t''_1 = 0$, $t'_1=t_1$, $t'_3=t_3$, $t''_2 = t_2$, $t''_4=t_4$. The overall number of sites is $N=50$, and the defect position is located at $N_d= 25$.
 %(a)(b)Case I: A chain with $t_1=1$, $t_2=0.1$, $t_3=1$, $t_4=0.55$, $t'_4 = t''_3=t'_2 = t''_1 = 0$, $t'_1=t_1$, $t'_3=t_3$, $t''_2 = t_2$, $t''_4=t_4$.
 %(a)(b) A chain with $t_1=1$, $t_2=0.6$, $t_3=1$, $t_4=0.75$, $t'_4 = t''_3=t'_2 = t''_1 = 0$, $t'_1=t_1$, $t'_3=t_3$, $t''_2 = t_2$, $t''_4=t_4$.
 (a) The black curve and the green dots represent the spectra corresponding to the defect-free system under PBC and OBC, respectively. The red curve and the blue stars represent the spectra of the system with a defect under PBC and OBC, respectively. The arrows give the direction of winding as the bloch vector $k$ varies from $-\pi/a$ to $\pi/a$. Blue stars $A$ and $B$ correspond to the eigenenergies 
 of the system with a defect under OBC $E_A=-1.4978+0.25i$ and $E_B=-0.6174+0.0396 i$, respectively. (b) Profiles of the eigenstates for the system with a defect under OBC.}
	\label{fig 2}
\end{figure}

Zeng and Yu recently investigated a single defect in a non-reciprocal HN-model-based system consisting of an array of $N$ on-chip nanomagnets, where $N$ is relatively small \cite{Zeng2023R}. In their work, the PBC spectra used to wind the skin states under open boundary conditions were calculated based on the defect-free periodic structure. Figure \ref{fig 2}(a) shows the eigenenergies of the system with the defect under OBC (blue stars) and of the corresponding defect-free system under PBC (black curve) with parameters $t_1=1$, $t_2=0.6$, $t_3=1$, $t_4=0.75$, $t'_1=t_1$, $t'_3=t_3$, $t''_2 = t_2$, $t''_4=t_4$, $N=50$, and $N_d=25$. 
A strong defect is caused by the boundary condition with $t'_4 = t''_3=t'_2 = t''_1 = 0$.
The introduction of long-range hopping results in the creation of multiple twisted loops in the spectra of the defect-free system under PBC (black curve) \cite{wang2021generating}. If a reference energy is in the interior of the small winding loop, the winding number is $-2$ [Fig. \ref{fig 2}(a)]. We observe that all eigenenergies of the system with the defect under OBC (blue stars) are enclosed by the spectra of the defect-free system under PBC (black curve). This suggests that the NHSE dominates over the defect effect, and all states 
of the system with the defect under OBC
are therefore expected to be asymmetrically skewed to one side of this non-Hermitian system.
However, we find that defect states with substantial amplitudes around the defect position survive from the dominance of the NHSE, while all other states are aggregated at the right boundary by the skin effect [Fig. \ref{fig 2}(b)]. This significant discrepancy is actually due to the existence of the defect.

\begin{figure}[h]
	\centering
	\includegraphics[width=6cm]{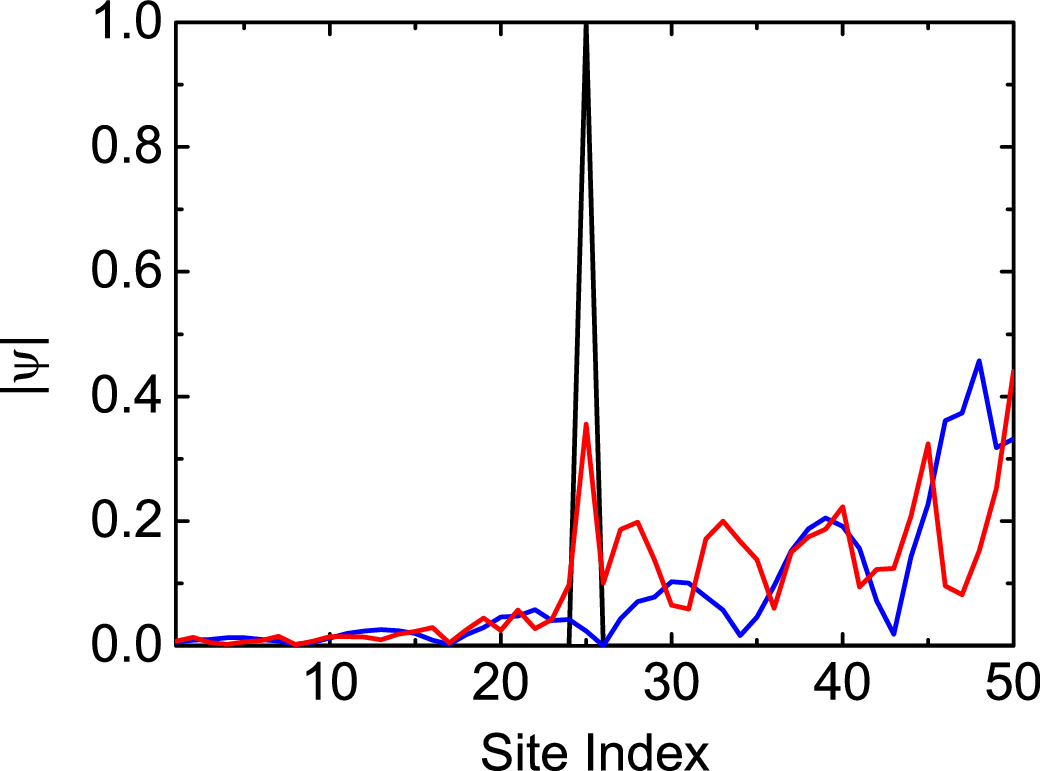}
 \caption{Profiles of the eigenstates corresponding to the eigenenergies of the system with a defect under OBC $E=0$ (black curve), $E_A=-1.4978+0.25i$ (blue curve), and
 $E_B=-0.6174+0.0396i$ (red curve) shown in Fig. \ref{fig 2}(a).}
	\label{fig 3}
\end{figure}

%To calculate the PBC spectrum for the system shown in Fig. \ref{fig 1}, which contains a defect in a unit cell with $N$-site lattice.
%with a defect can be treated as a unit cell.
%To explore the effect of the defect, we consider the system with the defect in its unit cell.
Figure \ref{fig 2}(a) also shows the PBC spectra of the system with the defect (red curve). These defect PBC spectra are quite different from the PBC spectra of the defect-free system (black curve).
We note that some purely real eigenenergies of the system with the defect under OBC (blue stars) are located on the section of the defect PBC spectra (red curve) which is on the real axis [Fig. \ref{fig 2}(a)]. Thus, their winding number is $0$.
%are exactly on the line with purely real energies of the red curve, which are enclosed by the red curve with zero winding number [Fig. \ref{fig 2}(a)]. 
This indicates that the corresponding states are defect states rather than skin states, and are thus not skinned at the boundary.
For example, the eigenenergy $E=0$ of the system with the defect under OBC is located on the red curve [Fig. \ref{fig 2}(a)], and the profile of the corresponding eigenstate confirms localization at the defect position (black curve in Fig. \ref{fig 3}). It is worth noting that the non-Hermitian point gap topology for coexistence of the NHSE and defects corresponds to a closed loop traced by the PBC spectra of the system which includes the defects in its unit cell, instead of the PBC spectra of the defect-free system.
%Topologically, as shown in Fig.\ref{Fig2}(b), this results in the OBC spectrum, except for the eigenenergy of the defect state, being enclosed by the spectrum corresponding to the perfect system under PBC, where the defect state caused by the boundary conditions ($t'_4 = t''_3=t'_2 = t''_1 = 0$) is not nontrivial.

To further unveil how the defect affects the system, we also show in Fig. \ref{fig 2}(a) the OBC spectra of the defect-free system (green dots). Comparing the OBC spectra with and without the defect elucidates how the defect-free OBC spectra (green dots) evolve to the defect OBC spectra (blue stars). A large section of the OBC spectra of the system with the defect matches well with the OBC spectra of the defect-free system [Fig. \ref{fig 2}(a)]. This suggests that the majority of the eigenstates are still dominated by the NHSE when only one defect is introduced into the system. Figure \ref{fig 3} verifies that the eigenstate with eigenenergy $E_A=-1.4978+0.25i$ 
in the defect OBC spectra
[marked by blue star $A$ in Fig. \ref{fig 2}(a)], which is 
very close to an eigenenergy in the defect-free OBC spectra (green dots),
%on the green-dot curve, 
rapidly decays with distance from the right boundary (blue curve in Fig. 3).
However, some states of the defect-free system under OBC evolve to defect states of the system with the defect under OBC, and their eigenenergies (blue stars) are thus located on the defect PBC spectra (red curve) [Fig. \ref{fig 2}(a)].
%the eigenenergy curve of the defect PBC spectra (red curve). 
%As shown in Fig. \ref{fig 2}(a), some defect OBC eigenenergies (blue stars) are located on the eigenenergy curve under PBC (red curve), and 
The number of times this curve winds around these defect OBC eigenenergies is zero.

Interestingly, we find that some eigenenergies of the defect-free system under OBC (green dots) evolve to eigenenergies of the system with the defect under OBC (blue stars) which are still encircled by the defect PBC spectra (red curve) but their positions largely deviate from the original defect-free OBC spectra [Fig. \ref{fig 2}(a)]. For example, the eigenstate corresponding to eigenenergy $E_B=-0.6174+0.0396 i$ 
in the defect OBC spectra
[blue star $B$ in Fig. \ref{fig 2}(a)] exhibits both localization around the defect and rapid decay away from the right boundary (red curve in Fig. \ref{fig 3}). Such eigenstates form a novel class of states which we refer to as hybrid skin-defect states. We found that such hybrid states result from the finite-size effect. In finite-size systems and especially in systems with small number of sites, skin states may extend to the defect site and couple to defect states to form hybrid states.  In contrast, in the thermodynamic limit, the skin and defect states are far away from each other and do not overlap. Thus, no hybrid states exist.

\begin{figure}[h]
\centering
\includegraphics[width=\columnwidth]{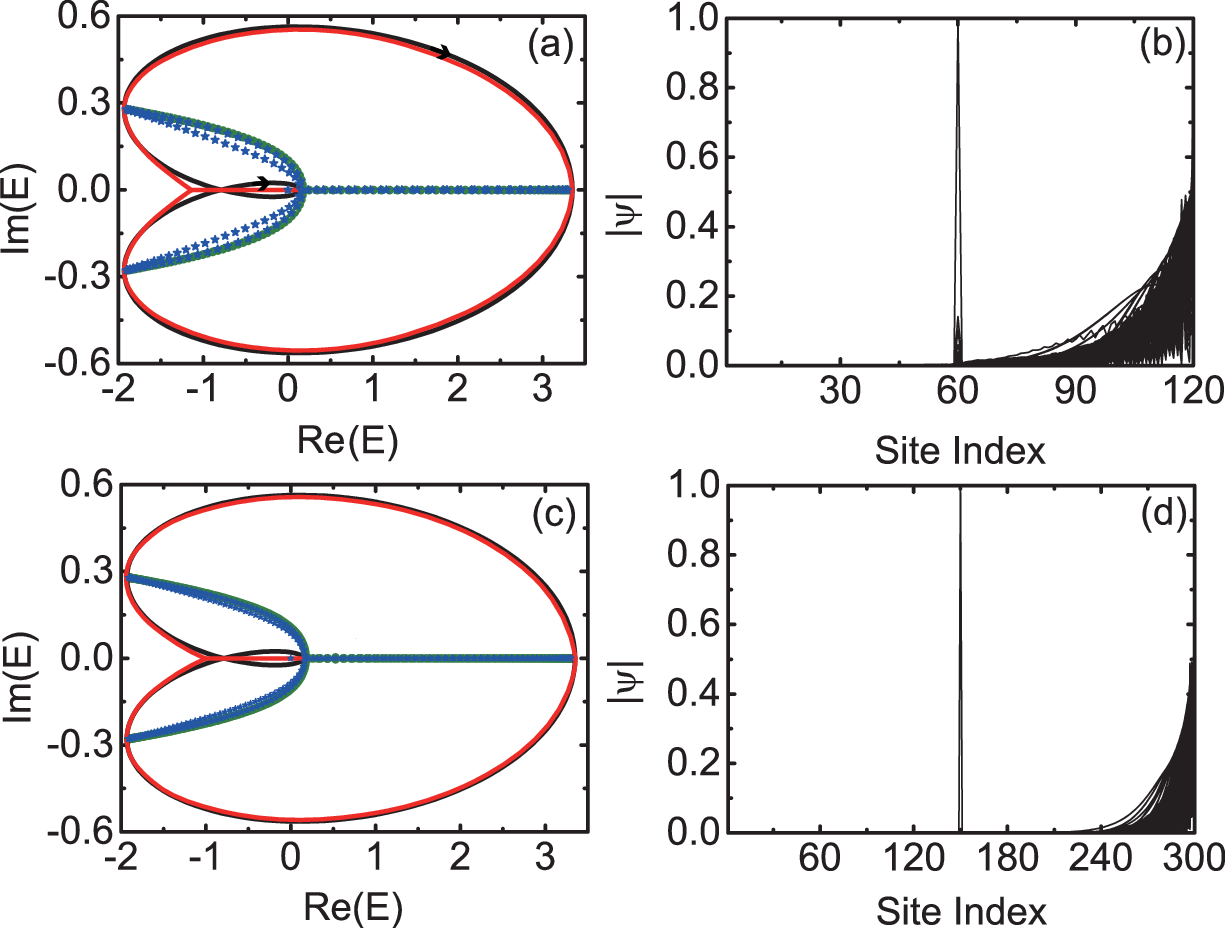}
\caption{The energy spectra of the one-dimensional non-Hermitian HN model with (a) $N=120$, $N_d=60$, and (c) $N=300$, $N_d=150$. The black, green-dot, red, and blue-star curves represent the spectra corresponding to the defect-free system under PBC and OBC, and the system with a defect under PBC and OBC, respectively. Profiles of the eigenstates of the system with a defect under OBC for (b) $N=120$, $N_d=60$, and (d) $N=300$, $N_d=150$. All other parameters are as in Fig. \ref{fig 2}.}
\label{fig 4}
\end{figure}

To further explore the size effect on the hybrid skin-defect states, we investigate systems with different sizes. Figure 4(a) shows the energy spectra of our proposed HN-model-based system for $N=120$ and $N_d=60$.
Compared to the case of $N=50$ and $N_d=25$ [Fig. \ref{fig 2}(a)], we observe that the defect OBC eigenenergies (blue stars) for the larger system with $N=120$, $N_d=60$ are closer to the defect-free OBC eigenenergies (green dots). In addition, the number of defect states is largely suppressed. Only one defect state, corresponding to $E=0$, caused by the boundary condition in the defect region remains. In Fig. \ref{fig 4}(b) we observe that the amplitudes of the hybrid skin-defect states at the defect position significantly decrease. As the lattice size increases, the coupling between the skin and defect states decreases. In the limit of infinitely long lattice, the defect OBC eigenenergies (blue stars) coincide with the defect-free OBC eigenenergies (green dots), except for the single defect state corresponding to $E=0$ [Fig. \ref{fig 4}(c)]. In other words, except for the $E=0$ state, all other states are dominated by the NHSE and are localized at one boundary [Fig. \ref{fig 4}(d)].

\begin{figure}[H]
	\centering
	\includegraphics[width=6cm]{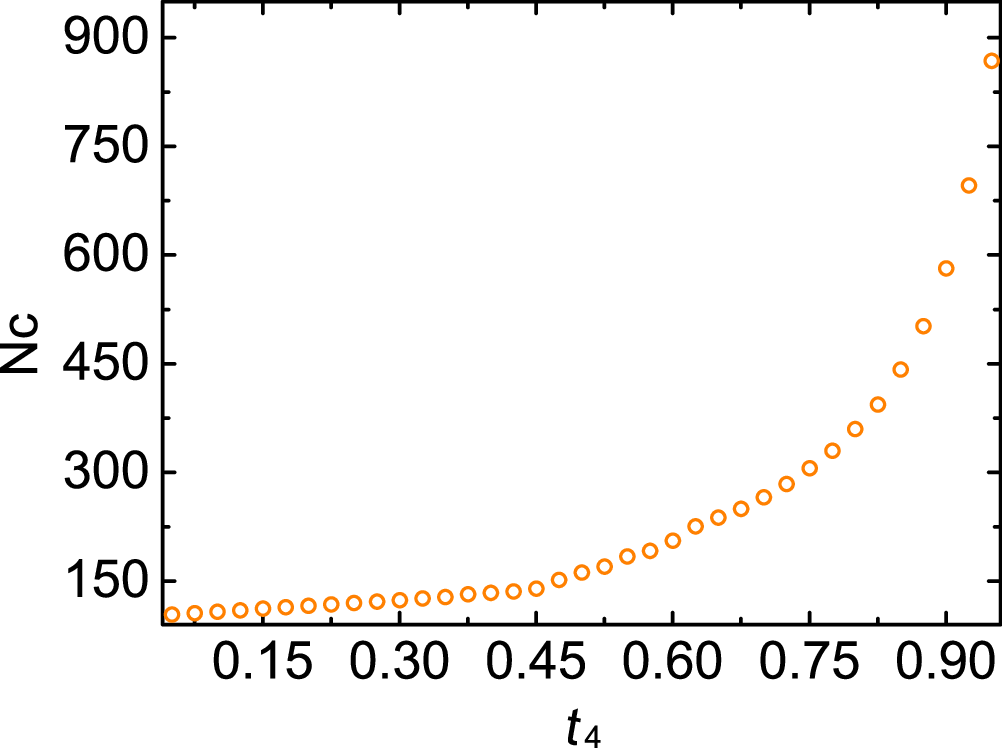}
	\caption{The critical value $N_c$ of the overall lattice size to completely suppress the hybrid skin-defect states as a function of the hopping term $t_4$. All other parameters are as in Fig. \ref{fig 2}.} %(b) The average of the eigenstates for the system with a defect at a random site under OBC. An extended state is shown as the red curve. All other parameters are as in Fig. \ref{fig 2}.}
\label{fig 5}
\end{figure}

Based on the analysis above, the system states go through a transition from hybrid states to skin states when the lattice size increases to a critical value $N_c$, after which all states, except for the zero-energy defect state, experience the NHSE and no hybrid states remain.
Figure \ref{fig 5}(a) shows the critical value $N_c$ as a function of the hopping term $t_4$. As $t_4$ increases, the non-reciprocity strength for NNN hopping $\left|1- t_4/t_3\right|$ decreases ($t_3=1$ in our model). We found that the critical value $N_c$ increases with $t_4$. For large $t_4$ the non-reciprocity strength is weak, the bulk states slowly decrease away from the boundary, and a large lattice size is therefore required to prevent the coupling of the bulk and defect states.

As shown in Figs. \ref{fig 2}(a), \ref{fig 4}(a) and \ref{fig 4}(c), the topology of the energy spectra of the defect-free system under PBC (black curves) is changed by a local and topologically trivial defect, and the small loop 
in the defect-free PBC spectra (black curves)
collapses into a straight line
in the defect PBC spectra (red curves). 
To exhibit the zero-energy defect state, the small loops in the energy spectra of the defect-free system under PBC (black curves) evolve to straight lines on the real axis in the energy spectra of the system with a defect under PBC (red curves).
%the defect PBC spectra [red  curves in Figs. \ref{fig 2}(a), \ref{fig 4}(a) and \ref{fig 4}(c)] must contain a curve in the real axis which is flattened from the small circle of the defect-free PBC spectra and is topologically trivial.
It is interesting that a local defect is able to, at least partially, change the point gap topological property of the system. We emphasize again the importance of the defect PBC spectra rather than the ones of its defect-free counterpart in the point gap topology of the structure with a defect. 

%In addition, Fig. \ref{fig 5}(b) shows the average of the eigenstates for 10000 configurations which include a defect at a random position. We observe that an extended state with an average of $\sim\frac{1}{N}$ appears (red curve), corresponding to the random defect states.

%\begin{figure}[htbp]
%    \centering
%\includegraphics[width=6.5cm]{fig 6}
%   \caption{The average eigenstates for the system with a defect at random site under OBC. A extended state is shown as the red curve. Other parameters are the same as in Fig.\ref{fig 2}.}
%    \label{fig 6}
%\end{figure}

\section{Single defect in a non-reciprocal SSH-model-based system}

\begin{figure}[h]
	\centering
	\includegraphics[width=\columnwidth]{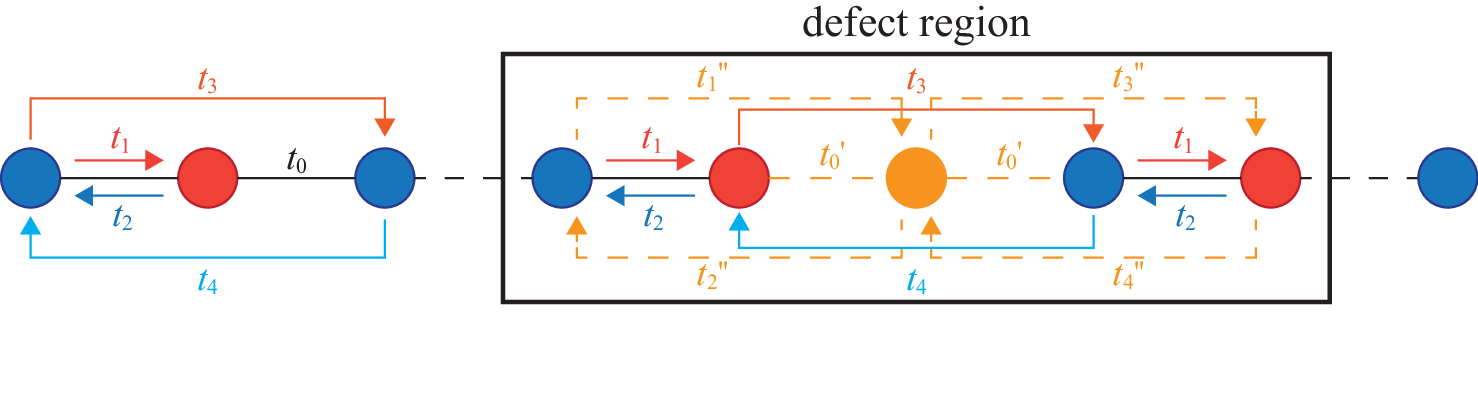}
	\caption{
    Schematic of a non-Hermitian one-dimensional SSH model with a defect.
    }
	\label{fig 6}
\end{figure}

We now extend our study to a non-reciprocal SSH-model-based system \cite{Su1979} with a single defect (Fig. \ref{fig 6}). Unlike the single-band HN-model-based system, discussed in the previous section, the non-reciprocal SSH-model-based system with NNN hoppings exhibits nontrivial band topology, making the defect problem more intriguing due to the coexistence of topologically trivial and nontrivial defect states. The Hamiltonian of this model is given by
\begin{equation} \label{modelssh}
H_{SSH}=H_{SSH}^L + H_{SSH}^R+H_{SSH}^d.
\end{equation}
The first two terms are the Hamiltonians of the left and right branches except the defect region:
\begin{align} \label{modelssh1}
H_{SSH}^{L(R)} &=\sum_{n=1}^{N_{L(R)}}(t_2c_{n,A}^{\dag}c_{n,B}+t_1c_{n,B}^{\dag}c_{n,A}) \\
&+ \sum_{n=1}^{N_{L(R)}-1}(t_0c_{n,B}^{\dag}c_{n+1,A}+t_0c_{n+1,A}^{\dag}c_{n,B}) \notag \\
%&+ \sum_{n=1}^{N_{L(R)}-1} (t_4c_{n,A}^{\dag}c_{n+1,A}+t_3c_{n+1,A}^{\dag}c_{n,A}) \notag\\
&+\sum_{\alpha=A,B}\sum_{n=1}^{N_{L(R)}-1}(t_4c_{n,\alpha}^{\dag}c_{n+1,\alpha}+t_3c_{n+1,\alpha}^{\dag}c_{n,\alpha}), \notag
\end{align}
where $c_{n,\alpha}$ ($c_{n,\alpha}^{\dag}$) is the annihilation (creation) operator at site $\alpha$ of the
$n^{\rm th}$ unit cell in each part of the SSH lattice (a unit cell contains two inequivalent sites $\alpha = A, B$); $N_{L(R)}$ is the number of unit cells in the left (right) branch; $t_1$ and $t_2$ are non-reciprocal intracell hopping terms; $t_3$ and $t_4$ are non-reciprocal NNN intercell hopping terms, and $t_0$ describes the reciprocal NN intercell hopping.

$H^d_{SSH}$ corresponds to the Hamiltonian of the defect region, which is given by
\begin{eqnarray} \label{modelssh2}
H_{SSH}^d & =& ({t'_0}c_{N_d-1,B}^{\dag}c_{N_d} +{t'_0}c_{N_d+1,A}^{\dag}c_{N_d}+h.c.)  \notag \\
&+& ({t''_2}c_{N_d-1,A}^{\dag}c_{N_d}+{t''_1}c_{N_d}^{\dag}c_{N_d-1,A}) \notag \\
&+& ({t''_3}c_{N_d+1,B}^{\dag}c_{N_d}+{t''_4}c_{N_d}^{\dag}c_{N_d+1,B}) \notag \\
&+& ({t_3}c_{N_d+1,A}^{\dag}c_{N_d-1,B}+{t_4}c_{N_d-1,B}^{\dag}c_{N_d+1,A}),
\end{eqnarray}
where $t'_0$ is the intercell hopping in the defect region, and $t''_1$, $t''_2$, $t''_3$, and $t''_4$ are non-reciprocal NNN hopping terms in the defect region; $N_d$ is the position of the single defect.

\begin{figure}[htbp]
	\centering
	\includegraphics[width=\columnwidth]{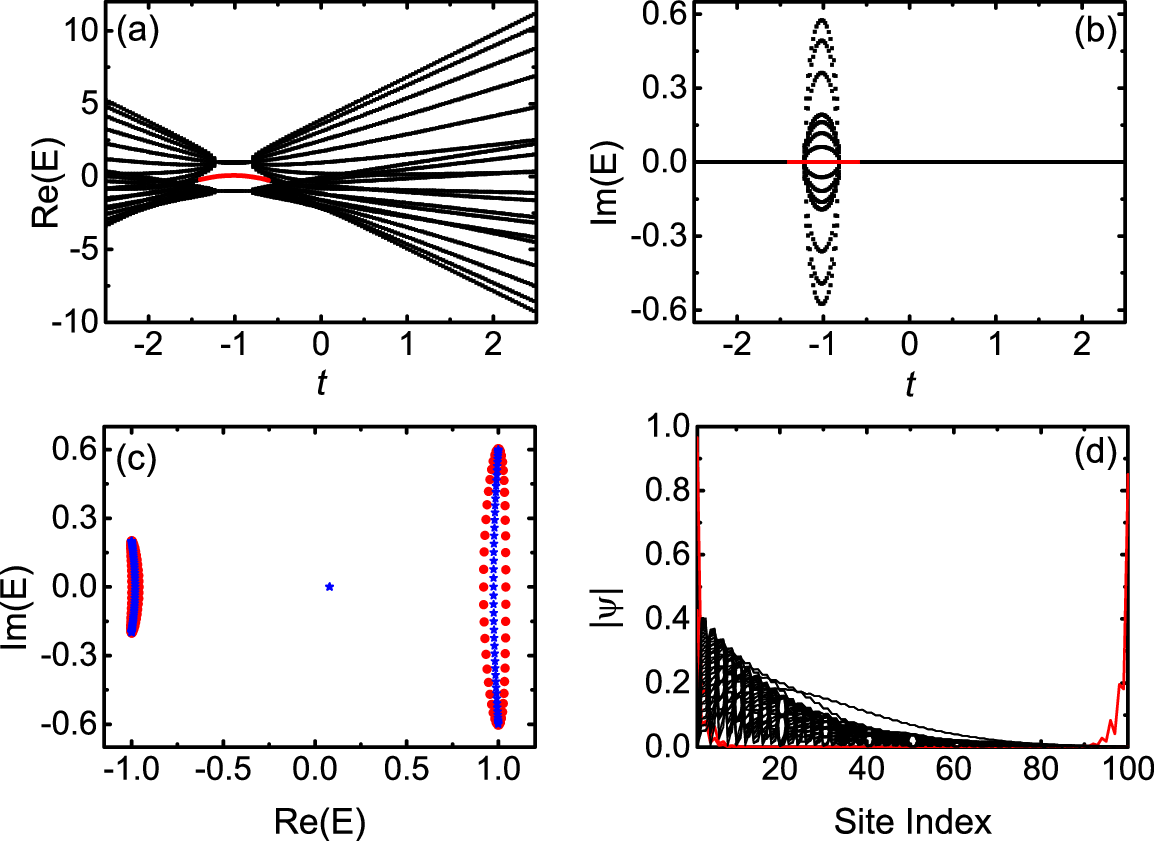}
	\caption{The topological properties of the non-Hermitian SSH chain without any defects. (a), (b) Real and imaginary parts of the eigenenergies under OBC of the non-Hermitian SSH-model-based defect-free system, with hopping parameters $t_1=t+e^\gamma$, $t_2=t+e^{-\gamma}$, and $\gamma = 0.2$. Other parameters are taken to be $t_3=t_1$, $t_4=t_2$, and $N=20$. Red (black) lines correspond to the topological edge (bulk) states. (c) The spectra of the non-Hermitian SSH-based defect-free system under PBC (red dots) and OBC (blue stars) for $t=-1$. (d) The eigenstates under OBC corresponding to the eigenenergies in (c). The profiles shown in red are topological edge states and correspond to the degenerate eigenenergies not enclosed by the PBC spectra. The profiles shown in black are skin states and correspond to the eigenenergies enclosed by the PBC spectra.}
	\label{fig 7}
\end{figure}

Before discussing the defect effect, it is important to investigate the topological properties of the non-Hermitian SSH-model-based lattice without any defect by examining symmetries. For systems with $t_1 \neq t_2$ and $t_3 \neq t_4$, only the time-reversal symmetry %($\mathcal{TRS}^\dagger$)
is satisfied as $\mathcal{T} H^{T} \mathcal{T}^{-1}=H$ \cite{kawabata2021}, %Operators $\mathcal{TRS}^\dagger$ is defined as
where $\mathcal{T}=I_N \otimes \sigma_y $ with $\mathcal{T}^2 = -1$, in which $\sigma_y$ is the $y$ component of the Pauli matrix, $N$ is the number of unit cells in the SSH chain, and $I_N$ is the $N \times N$ unit matrix. Thus, in this case the system belongs to the $AII^\dagger$ class and has the $Z_2$ class topological invariant $\nu \in \{0, 1\}$ \cite{PhysRevX.9.041015}. Even in the presence of non-reciprocal terms ($t_1 \neq t_2$, $t_3 \neq t_4$, and $t_{1,2}= t+e^{\pm \gamma}$), the system undergoes the process of gap closure and reopening with the increase of $t$. %[Figs. \ref{fig 7} (a) and (b)].
As shown in Figs. \ref{fig 7}(a) and \ref{fig 7}(b), the energy gap is closed and then open by the topological property of the system for $\gamma=0.2$ between $t= -1.4$ and $t = -0.6$. Meanwhile, the topological edge states come up between these two critical points [red line in Fig. \ref{fig 7}(a)]. When the system parameter $t \in [-1.2,-0.8]$, the imaginary part of the energy spectra is non-zero, indicating that non-reciprocal hoppings induce complex eigenvalues in the bulk states. In contrast, the topological edge states always have purely real eigenvalues [red line in Fig. \ref{fig 7}(b)].
In addition, under OBC, both the topologically trivial bulk states and topological edge states localize at the boundary. The skin localization of bulk states occurs because the two detached bulk spectra under OBC are encircled by the spectra under PBC, exhibiting the NHSE. The topological edge states correspond to the two degenerate eigenenergies which lie in the line gap [Fig. \ref{fig 7}(c)]. Figure \ref{fig 7}(d) shows that the two degenerate edge states (red lines) are localized at both sides of the system, while all other states (black lines) are aggregated at the left boundary.

\begin{figure}[]
\centering
\includegraphics[width=\columnwidth]{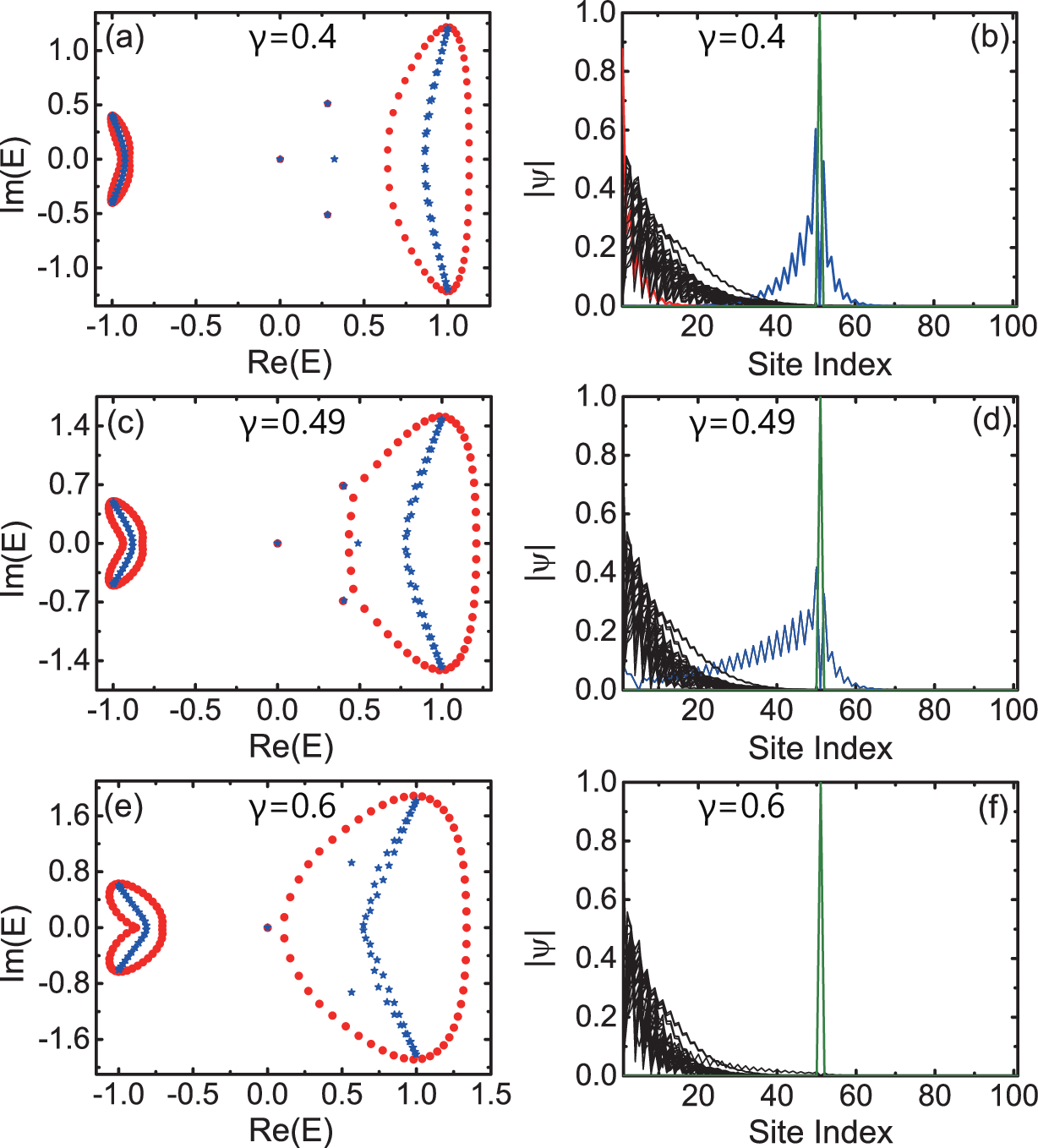}
\caption{The evolution of the eigenenergies and eigenstate profiles of the non-Hermitian SSH system with a defect shown in Fig. \ref{fig 6}. The hopping parameters are $t_1=t+e^\gamma$, $t_2=t+e^{-\gamma}$, $t=-1$, $t_3=t_1$, $t_4=t_2$, $t_0'=t_{1,2,3,4}''=0$, with varying $\gamma$. The overall number of sites is $N=101$, the defect is located at the site $N_d=51$, and the number of SSH unit cells is $50$. The energy spectra under PBC (red dots) and OBC (blue stars) are shown for (a) $\gamma=0.4$, (c) $\gamma=0.49$, and (e) $\gamma=0.6$. The corresponding state profiles for $\gamma=0.4$, $\gamma=0.49$, and $\gamma=0.6$ under OBC are shown in (b), (d), and (f), respectively. Red, blue, green, and black lines represent the topological edge states, the topologically nontrivial defect states, the trivial defect state, and the skin states, respectively.}
\label{fig 8}
\end{figure}

We now introduce a strong defect into our proposed SSH-based system (Fig. \ref{fig 6}), that is $t_0'=t_{1,2,3,4}''=0$. We explore how this system evolves with $t_1=t+e^\gamma$, $t_2=t+e^{-\gamma}$, $t_3=t_1$, $t_4=t_2$, and $t=-1$ for different $\gamma$. For the system with PBC, the overall lattice size of the chain is $N=101$ and the defect is located at $N_d=51$, so that there are $25$ SSH unit cells to the left and to the right of the defect. When $\gamma$ increases from 0.2 to 0.4, five eigenenergies appear in the line gap, with two of them being degenerate [Fig. \ref{fig 8}(a)]. Among them, the eigenenergy $E=0$, also appearing on the defect PBC spectra, corresponds to the trivial defect state caused by the boundary condition in the defect region [green line in Fig. \ref{fig 8}(b)]. Two conjugate eigenenergies, also appearing on the defect PBC spectra [Fig. \ref{fig 8}(a)], correspond to the nontrivial $Z_2$ topological states around the defect. We observe that these two states with conjugate eigenenergies in fact share the same eigenstate profile [blue line in Fig. \ref{fig 8}(b)]. In addition, two degenerate eigenenergies under OBC, which are not encircled by the defect PBC spectra [Fig. \ref{fig 8}(a)], correspond to the $Z_2$ topological edge states on the open boundary [red line in Fig. \ref{fig 8}(b)]. In contrast to the case of $\gamma=0.2$ shown in Figs. \ref{fig 7}(c) and \ref{fig 7}(d), the stronger NHSE induces a coupling between the left and right topological edge states and forces the right edge state to localize towards the left boundary, competing with the topological protection of the time-reversal symmetry \cite{Cheng2022Competition}. Although both $Z_2$ edge states localize at the left boundary, they are topologically different from the other skin states. All other eigenenergies obtained under OBC which are encircled by the defect PBC spectra [Fig. \ref{fig 8}(a)], correspond to skin states [black lines in Fig. \ref{fig 8}(b)].
	
As $\gamma$ increases to $\gamma=0.49$ [Fig. \ref{fig 8}(c)], the NHSE becomes stronger, and the area enclosed by the defect PBC spectra expands. As $\gamma$ increases, except for the eigenenergy of $E=0$ caused by the boundary condition in the defect region, the other eigenenergies in the line gap gradually move towards the defect PBC spectra. Fig. \ref{fig 8}(c) shows that, when $\gamma=0.49$, two degenerate eigenenergies under OBC corresponding to topological edge states are enclosed by the defect PBC spectra, and thus lose their competition to the NHSE. Fig. \ref{fig 8}(d) shows the corresponding state profiles for $\gamma=0.49$. Although the two complex conjugate eigenenergies in the line gap survive from the NHSE for $\gamma=0.49$, as $\gamma$ further increases to $\gamma=0.6$, they finally fall into the region enclosed by the defect PBC spectra and transit to skin states [Fig. \ref{fig 8}(e)]. Fig. \ref{fig 8}(f) shows the corresponding state profiles for $\gamma=0.6$, where all the states become topologically trivial due to the strong NHSE.
%The topological nontrivial defect states formed between the SSH chain and the defect are deeply suppressed. Exception of the trivial defect state corresponding to $E=0$, all other states are dominated by the NHSE.
This indicates that the trivial defect states are insensitive to the NHSE if the defect is strong enough, while the topologically nontrivial defect states cannot compete with the NHSE if the non-reciprocity is strong enough. When $\gamma$ is small ($\gamma=0.2$), the NHSE is too weak to compete with the topological protection of the time-reversal
symmetry. The topological edge states locate at the two open boundaries [Fig. \ref{fig 7}(d)]. As $\gamma$ increases ($\gamma=0.4$), one edge state shifts to the other side at which the skin states localize due to the NHSE enhancement [Fig. \ref{fig 8}(b)]. If $\gamma$ further increases ($\gamma=0.49$), the two degenerate topological edge states lose the competition to the NHSE and become skin states [Fig. \ref{fig 8}(d)]. Finally, when $\gamma$ is large enough ($\gamma=0.6$), the NHSE dominates the topological protection of the time-reversal symmetry and the nontrivial defect states evolve to skin states as well [Fig. \ref{fig 8}(f)].

\begin{figure}[]
	\centering
	\includegraphics[width=\columnwidth]{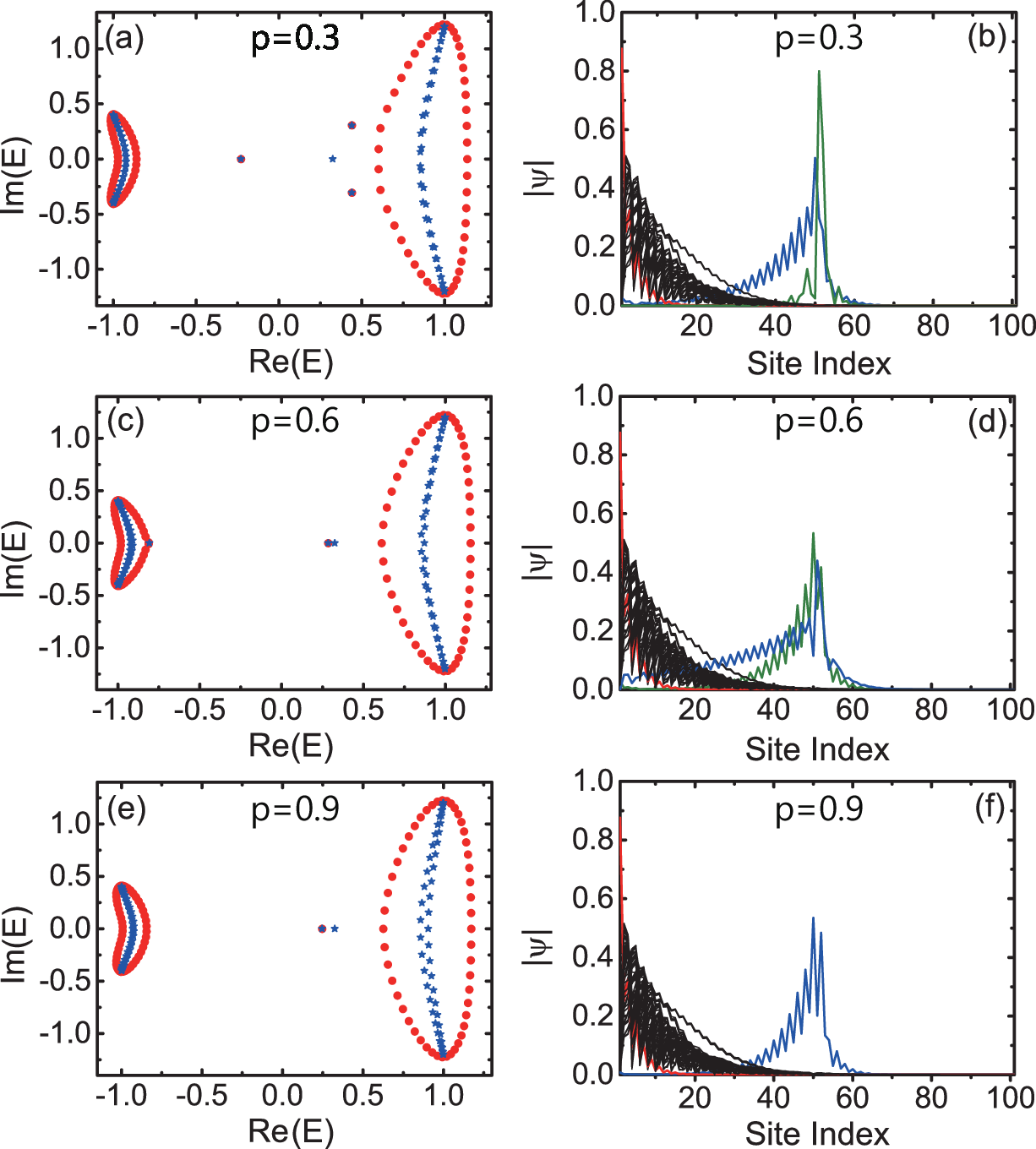}
	\caption{The effect of the defect strength in the presence of the NHSE. The hopping parameters are $t_1=t+e^\gamma$, $t_2=t+e^{-\gamma}$, $t=-1$, $\gamma=0.4$, $t_3=t_1$, $t_4=t_2$, $t_1''=t_3''=p t_3$, $t_2''=t_4''=p t_4$, $t_0=pt_0'$, with varying $p$.
    The overall number of sites is $N=101$, the defect is located at the site $N_d=51$, and the number of unit cells in the SSH chain is $50$.
    The energy spectra under PBC (red dots) and OBC (blue stars) are shown for (a) $p=0.3$, (c) $p=0.6$, and (e) $p=0.9$. The eigenstate profiles under OBC corresponding to (a), (c), and (e), are shown in (b), (d), and (f), respectively. Red, blue, green, and black lines represent the topological edge states, the topologically nontrivial defect states, the trivial defect state, and the skin states, respectively.}
	\label{fig 9}
\end{figure}

Next, we consider how the boundary condition in the defect region affects the defect states in the presence of the NHSE. More specifically, we consider the effect of $p$, which is a parameter controlling the hopping in the defect region through $t_1''=t_3''=p t_3$, $t_2''=t_4''=pt_4$, and $t_0'=pt_0$. As $p$ increases from $0$ to $1$, the strength of the defect decreases from strong to weak. Compared to the case of $p=0$ [Fig. \ref{fig 8}(a)], the eigenenergy for the trivial defect state shifts from $E=0$ towards the left branch of the PBC spectra for $p=0.3$ [Fig. \ref{fig 9}(a)], and the corresponding trivial defect state shrinks and broadens [green line in Fig. \ref{fig 9}(b)]. When $p$ is increased to $0.6$, the eigenenergy of the trivial defect state almost reaches the boundary of the left PBC spectra [Fig. \ref{fig 9}(c)], and the corresponding trivial defect state further shrinks and broadens [green line in Fig. \ref{fig 9}(d)]. In addition, the topologically protected pair of conjugate eigenenergies becomes real and degenerate [Fig. \ref{fig 9}(c)]. When $p$ is increased to 0.9, the strength of the defect further decreases. The eigenenergy which corresponded to the trivial defect state for smaller values of $p$ is now enclosed by the left PBC spectra [Fig. \ref{fig 9}(e)]. Thus, for $p=0.9$ no trivial defect state exists 
%and this state is completely suppressed and disappears 
[Fig. \ref{fig 9}(f)]. This shows that the defect has to be strong enough for the trivial defect state to survive in the competition with the NHSE. In contrast, the topologically nontrivial defect states [blue curves in Figs. \ref{fig 9}(b), \ref{fig 9}(d), and \ref{fig 9}(f)] exist for all values of $p$ due to the robustness of the topological protection. It is noteworthy that the bulk states in our proposed SSH-model-based system are topologically nontrivial. As a result, compared to the HN model that we considered in the previous section, here there is no coupling between the skin states and trivial defect states. Thus, in our proposed SSH-model-based system we do not observe finite-size effects and hybrid skin-defect states similar to the ones in our proposed HN-model-based system.

\section{Discussion and conclusions}
In summary, we first investigated the interplay of the NHSE and a defect in non-reciprocal HN lattices. We demonstrated that the non-Hermitian point gap topology for coexistence of the NHSE and defects corresponds to a closed loop which should be traced by the PBC spectra of the system with unit cell including the defects in the complex plane. A local defect is capable of changing the point gap topological property of the system. We observed a novel class of hybrid skin-defect states in finite-size systems which originate from the coupling between the skin and defect states. A transition from hybrid skin-defect states to skin states occurs when the coupling between the defect and the skin states is varied, and depends on the size and the non-reciprocity strength of the system. We also investigated the defect effect in the topologically nontrivial %lattice with time-reversal symmetry based on the non-reciprocal
SSH model. We showed that increasing the non-reciprocity strength can lead to a transition from topologically nontrivial defect states to skin states, while the trivial defect state is insensitive to the NHSE, if the strength of the defect is large enough. In addition, decreasing the defect strength can result in a transition from trivial defect states to skin states.
By tuning the position of the defect, the topologically nontrivial defect states can also be shifted accordingly.
Our work reveals the fundamental interplay between defects and non-Hermiticity, which can serve as a universal mechanism for non-Hermitian topological physics. The results can also be generalized to higher dimensions. Our models and results can be implemented experimentally in systems with non-reciprocal couplings \cite{weidemann2020topological,wang2021generating,zhang2021acoustic}.

Y.H. was supported by the Natural Science Foundation of Hunan Province Grant No. 2024JJ5426m, and W.L. was supported by the NSF-China under Grant No. 11804396. This work was supported in part by the High Performance Computing Center of Central South University.
	
\twocolumngrid
\section*{Reference}
\makeatletter
\renewcommand{\bibsection}{}
\makeatother
\bibliography{ref}
\end{document}